\documentclass[a4paper,fleqn]{cas-dc}

\usepackage[authoryear]{natbib}

\usepackage[linesnumbered,ruled,vlined]{algorithm2e}
\usepackage[utf8]{inputenc}
\usepackage{subcaption} 

\usepackage{amsmath} 
\def\tsc#1{\csdef{#1}{\textsc{\lowercase{#1}}\xspace}}
\tsc{WGM}
\tsc{QE}
\tsc{EP}
\tsc{PMS}
\tsc{BEC}
\tsc{DE}

\begin{document}
\let\WriteBookmarks\relax
\def\floatpagepagefraction{1}
\def\textpagefraction{.001}

\fontsize{8pt}{\baselineskip}\selectfont  

\shorttitle{}
\shortauthors{Y. Cheng et~al.}

\title [mode = title]{Robust Steganography with Boundary-Preserving Overflow Alleviation and Adaptive Error Correction}                      

\tnotetext[1]{This research work was supported by the National Natural Science Foundation of China No.62172001.}

\author[1]{Yu Cheng}[orcid=0009-0005-3128-5613
                        ]

\ead{10212140443@stu.ecnu.edu.cn}

\affiliation[1]{organization={Shanghai Key Laboratory of Multidimensional Information Processing},
                addressline={East China Normal University}, 
                city={Shanghai},
                postcode={200241}, 
                country={China}}

\author[1]{Zhenlin Luo}[orcid=0009-0001-3596-8009
                        ]

\ead{10212140438@stu.ecnu.edu.cn}


\author[1]{Zhaoxia Yin}[orcid=0009-0003-0387-4806
                       ]
\cormark[1] 
\ead{zxyin@cee.ecnu.edu.cn}


\cortext[cor1]{Corresponding author}

\begin{abstract}
With the rapid evolution of the Internet, the vast amount of data has created opportunities for fostering the development of steganographic techniques. However, traditional steganographic techniques encounter challenges due to distortions in online social networks, such as JPEG recompression. Presently, research into the lossy operations of spatial truncation in JPEG recompression remains limited. Existing methods aim to ensure the stability of the quantized coefficients by reducing the effects of spatial truncation. Nevertheless, these approaches may induce notable alterations to image pixels, potentially compromising anti-steganalysis performance. In this study, we analyzed the overflow characteristics of spatial blocks and observed that pixel values at the boundaries of spatial blocks are more prone to overflow. Building upon this observation, we proposed a preprocessing method that performs overflow removal operations based on the actual overflow conditions of spatial blocks. After preprocessing, our algorithm enhances coefficient stability while minimizing modifications to spatial block boundaries, favoring image quality preservation. Subsequently, we employed adaptive error correction coding to reduce coding redundancy, thereby augmenting robustness and mitigating its impact on anti-steganalysis performance. The experimental results indicate that the proposed method possesses a strong embedding capacity, maintaining a high level of robustness while enhancing security.
\end{abstract}

\begin{graphicalabstract}
\includegraphics[width=1.0\linewidth]{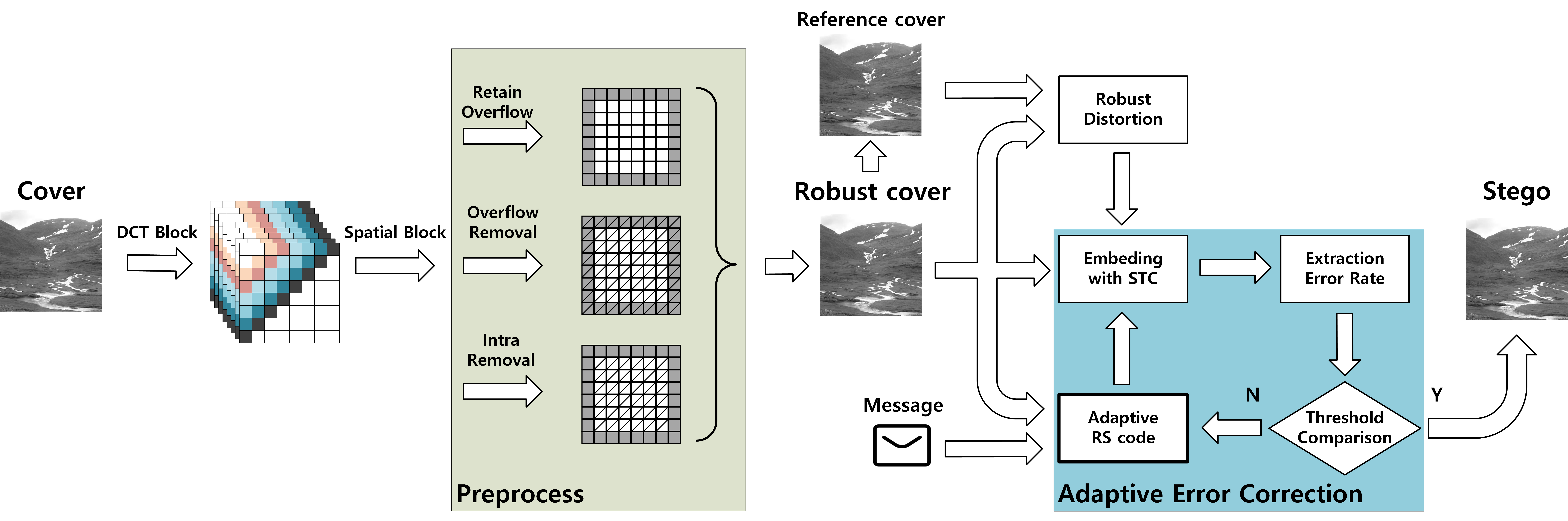}
\end{graphicalabstract}

\begin{highlights}
\item Spatial block boundaries are more prone to overflow in JPEG recompression.
\item Preprocessing minimizes block modifications, enhancing stability and performance.
\item Adaptive error correction promptly saves the optimal scheme to enhance performance.
\item Method demonstrates significant advantages in anti-steganalysis performance.
\end{highlights}


\maketitle

\section{Introduction}

{W}{ith} the rapid develpment of the Internet, images become the primary medium for exchanging information and the security of data has attracted increasing attention.  Data hiding \citep{XIAO2024104291, ZHANG2024124338, KONG2024124903, ZHANG2024120225} and encryption \citep{LI2024124574} are two major techniques currently used for information protection. Steganography \citep{HU2024123715, YAO2024123540, REHMAN2024123420, GAN2024121955, 10365238} as an crucial data hiding technique, involves embedding secret information into carriers to achieve covert communication, thereby effectively safeguarding data due to its unique concealment properties. To detect steganography, steganalysis \citep{GUO2024124796, holub2014low} has been proposed to determine whether an image has undergone steganographic embedding. The anti-steganalysis performance is an important metric for evaluating the security of a steganographic algorithm.

During online social networks (OSNs), such as Facebook and Twitter, JPEG images are extensively utilized due to their high compression efficiency and practicality. Numerous digital image steganographic techniques are designed for this format \citep{SONG2024124390,rustad2023digital,yin2021robust, SAJEDI20107703}. However, the lossy operations such as JPEG recompression, filtering, and scaling during the transmission of images in OSNs may lead to a decrease in the accuracy of extracting hidden information.  Especially due to the lossy operations involved in JPEG recompression, such as cofficients quantization, spatial truncation and spatial rounding operations, higher demands are placed on the robustness of steganographic techniques. 

Existing steganographic techniques resistant to JPEG compression can be categorized into three types \citep{yu2020robust}: 1) Upward Robust; 2) Matching Robust; 3) Downward Robust. Among them, Upward Robust methods refer to situations where the quality factor (QF) of the cover ($Q_{cover}$) is smaller than the QF of the channel ($Q_{channel}$). The practical significance of upward robust techniques is notable because users can manage the $Q_{cover}$. Additionally, they offer the advantages of behaving normally and requiring low computational complexity. Matching Robust refers to algorithms that perform well only when the $Q_{channel}$ before recompression is known. This category of methods is represented by TCM (Transport Channel Matching), proposed by Zhao et al. \citep{zhao2018improving}. Downward Robust refers to scenarios where the $Q_{channel}$ is lower than the $Q_{cover}$. This category is exemplified by image steganographic schemes proposed by Tao et al. \citep{tao2018towards}, who utilize mathematical derivations to achieve robustness. Additionally, Zeng et al. \citep{10057024} introduce a novel approach named PMAS (Postprocessing and Precise Dither Modulation based Robust Adaptive Steganography) specifically designed for high-quality images. However, both Matching Robust and Downward Robust methods have security vulnerabilities, enabling adversaries to exploit these weaknesses for steganalysis. These methods require uploading and downloading the original cover JPEG image once or several times from the specific OSN, which is a suspicious behavior.

Due to the avoidance of the issues encountered by the other two methods in practical applications, numerous techniques are specifically designed for upward robust methods \citep{yu2020robust, zhang2018dither, duan2023robust, 10093140}. Zhang et al. \citep{zhang2018dither} proposed a robust adaptive steganography called Dither Modulation Based Robust Adaptive Steganography (DMAS), aiming to achieve undetectability and robustness. This approach embeds information via dither modulation on mid-frequency alternating current (AC) coefficients. Yu et al. \citep{yu2020robust} extended dither modulation to generalized dither modulation and introduced asymmetric distortion in adaptive steganography based on generalized dither modulation (GMAS). Duan et al. \citep{duan2023robust} proposed a robust image steganographic algorithm leveraging image characteristics through embedding domain selection and adaptive error correction (Adaptive-GMAS). This algorithm adjusts the embedding domain and Reed-Solomon (RS) \citep{macwilliams1977theory} error correction capability based on the characteristics of different images, thus achieving a balance between robustness and security.

Currently, there is limited research on the lossy operation of spatial truncation in JPEG recompression. To address this issue, Zeng et al. \citep{10093140} proposed the Removal of Overflow-based Adaptive Robust Steganography Technique (ROAST). This method ensures the stability of quantized DCT coefficients by preprocessing the spatial pixel values. After preprocessing, the entire image can serve as a robust region, thus offering a larger embedding capacity while maintaining robustness. However, due to the potential for noticeable changes in image pixel values with this approach, it may have implications for the security of the algorithm.

To further mitigate the impact of spatial truncation on the performance of steganographic algorithms, we conducted a statistical analysis of spatial overflow in JPEG images and found that spatial block boundaries are more prone to overflow. Based on this observation, we developed a preprocessing method that selectively addresses overflow based on the actual overflow situation of spatial blocks. This approach reduces modifications to block boundaries, preserving inter-block correlations, thereby enhancing empirical security. Additionally, recognizing that fixed error correction coding may lead to coding redundancy or insufficient error correction capability, we adopted adaptive error correction coding in our method. This not only ensures robustness but also improves anti-steganalysis performance. The anti-steganalysis performance is evaluated by using CCPEV (PEV \citep{pevny2007merging} features enhanced by Cartesian calibration) \citep{kodovsky2009calibration}, DCTR \citep{holub2014low} and SRNet \citep{8470101}. Experimental results indicate that our steganographic technique exhibits a certain degree of robustness and excellent security.

The main contributions are summarized as follows:
\begin{itemize}
  \item [(1)]
In response to spatial truncation, we found that spatial block boundaries are more prone to overflow. Consequently, we designed a preprocessing method that selectively handles overflow according to the specific overflow situation of spatial blocks. This preprocessing method reduces modifications to block boundaries while maintaining the stability of quantized DCT coefficients, thereby improving the performance of the steganographic technique.
  \item [(2)]
The proposed method employs adaptive error correction coding to enhance robustness and reduce the impact on anti-steganalysis performance. During the iterative process of adaptive error correction, the error correction code corresponding to the currently optimal robustness is promptly saved to reduce redundancy in error correction capabilities.
  \item [(3)]
Compared to existing methods, the proposed method is an upward robust steganographic technique with a high embedding capacity. A wealth of experimental results demonstrates the significant advantage of the proposed method in anti-steganalysis performance, thereby highlighting its considerable utility in social network transmissions.
  
\end{itemize}

The remainder of this paper is organized as follows. Section \ref{section:Related Work} introduces symbol notation and related work. Section \ref{section:Proposed method} describes the methods employed in this study. Section \ref{section:Experiment} outlines the experiments and their results analysis. Finally, Section \ref{section:Conclusion} concludes the paper.

\section{Related Work}
\label{section:Related Work}
In this section, we introduce symbol notation, the JPEG recompression process, the dither modulation method, as well as the classical GMAS \citep{yu2020robust} algorithm and the advanced ROAST \citep{10093140} algorithm, aimed at enhancing comprehension of our method.

\subsection{Notations and JPEG Recompression}
Table \ref{tab:notation} summarises frequently used symbols in this paper. Matrices and vectors are represented in bold typeface, while individual elements within the vectors are indicated with unbolded lowercase letters.

\begin{table}
\centering
\renewcommand\arraystretch{1.5}
\scriptsize 
\caption{Notations}
\label{tab:notation}
\begin{tabular}{l | l}
\hline
\multicolumn{1}{c|}{Symbol} & Definition \\
\hline
 \multicolumn{1}{c|}{\textit{\textbf{D}}}& The quantized DCT coefficients of cover image \\
\multicolumn{1}{c|}{$\overline{\textit{\textbf{D}}}$}& The dequantized DCT coefficients of cover image \\
 \multicolumn{1}{c|}{\textit{\textbf{S}}}& The spatial values of cover image \\
 \multicolumn{1}{c|}{\textbf{\textit{Q}}}& Quantization table with 8 × 8 size \\
 \multicolumn{1}{c|}{\textbf{\textit{I}}}& The interior of the DCT block in spatial domain \\
 \multicolumn{1}{c|}{\textbf{\textit{C}}}&The corner of the DCT block in spatial domain\\
 \multicolumn{1}{c|}{\textbf{\textit{B}}}& The boundary of the DCT block in spatial domain \\
 \multicolumn{1}{c|}{\textit{m}}& Secret 
 binary message \\
 \multicolumn{1}{c|}{\textit{T1}}& Intensity parameter for overflow removal \\
 \multicolumn{1}{c|}{\textit{$\mu$}}& Parameter used to adjust asymmetric distortion \\
\hline
\end{tabular}
\end{table}

\begin{figure}[b]
    \centering
    \includegraphics[width=0.82\linewidth]{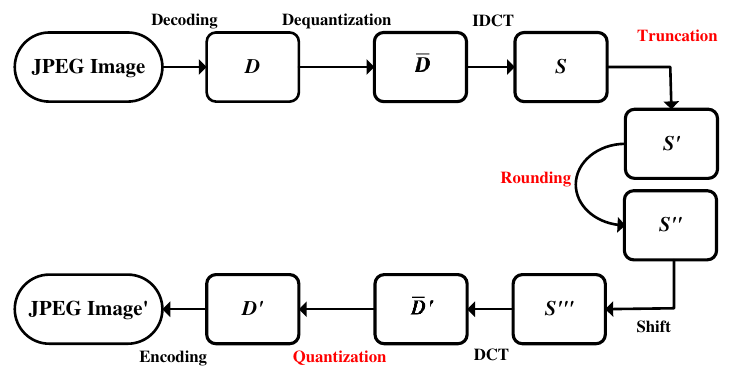}
    \caption{The process of JPEG recompression.}
    \label{fig:recompression}
\end{figure}

Fig. \ref{fig:recompression} illustrates the process of JPEG recompression. The process includes decoding the JPEG image to convert the DCT coefficients to spatial values using an IDCT transformation. The spatial domain image is then obtained through spatial truncation, spatial rounding and spatial shift. Subsequently, the spatial domain image undergoes DCT transformation, and the coefficients are quantized to quantized DCT coefficients with a JPEG compression $Q_{channel}$ according to the channel quantization table.

From the above description of the JPEG recompression process, we can easily identify its main lossy operations: spatial truncation, spatial rounding and coefficient quantization. The related equations are shown as formulas \eqref{con:trunc}, \eqref{con:round}, and \eqref{con:quant}. The notation [·] indicates rounding operation.

\begin{itemize}
    \item \textbf{spatial truncation:}
\end{itemize}
\begin{equation}
\label{con:trunc}
\mathbf{\textit{S}}^{\prime}=\mathbf{TRU}(\mathbf{\textit{S}})
\end{equation}

\begin{itemize}
    \item \textbf{spatial rounding:}
\end{itemize}
\begin{equation}
\label{con:round}
\mathbf{\textit{S}}^{\prime\prime}=[\mathbf{\textit{S}}^{\prime}+128]
\end{equation}

\begin{itemize}
    \item \textbf{coefficients quantization:}
\end{itemize}
\begin{equation}
\label{con:quant}
\mathbf{\textit{D}}^{\prime}=[\overline{\mathbf{\textit{D}}}^{\prime}/\mathbf{\textit{Q}}]
\end{equation}

In formula \eqref{con:quant}, \textbf{\textit{Q}} refers to the quantization table corresponding to the $Q_{channel}$. The function \textbf{TRU}(·) truncates the pixel value that falls outside the spatial range, as shown in formula \eqref{con:TRU}, where \textit{s} represents the spatial value. 

\begin{equation}
\label{con:TRU}
\mathbf{TRU}(s)=\begin{cases}127&s>127\\-128&s<-128\\s&else\end{cases}
\end{equation}

\begin{figure}[b]
    \centering
    \includegraphics[width=0.7\linewidth]{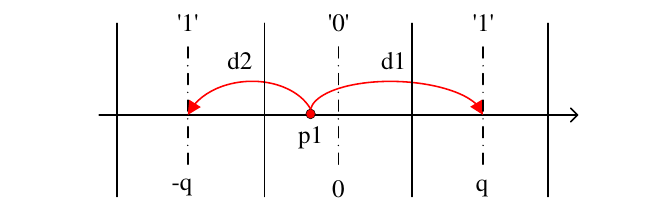}
    \caption{Dither modulation algorithm.}
    \label{fig:dither}
\end{figure}

\subsection{Dither Modulation}

Dither modulation is an implementation of the Quantization Index Modulation (QIM) scheme \citep{chen2001quantization, noda2006high}. As shown in Fig. \ref{fig:dither}, the quantization step from its quantization table divides the dequantized DCT coefficient into alternating odd and even intervals. During odd intervals, the cover element is represented as '1', while during even intervals, it is represented as '0'. The utilization of parity guarantees that the message bits are embedded without affecting the cover image. Additionally, the minimal modification distance is used to adjust the dequantized coefficient as illustrated in the Fig. \ref{fig:dither}. The value of \textit{p1} falls within the even interval thus corresponds to the information '0'. If \textit{p1} is required to represent the information '1', subtract the modified distance \textit{d2} from the current quantized coefficient based on the principle of minimum modified distance. Zeng et al. \citep {10093140} highlighted that dither modulation enhances the robustness of upward robust algorithms, making the embedding process error-free when considering only quantization and neglecting other lossy processes.

\subsection{GMAS}
\label{section:GMAS}

The GMAS \citep{yu2020robust} method employs generalized dither modulation and expands the embedding domain of DMAS \citep{zhang2018dither} to enhance its resistance to JPEG recompression operations and improve detection resistance. The embedding algorithm is briefly outlined as follows:

\begin{itemize}
  \item [1)]
Calculate the dequantized DCT coefficients for a given cover image. Evaluate the distortion $\rho_{ij}$ of each cover element using the J-UNIWARD \citep{holub2014universal} distortion function. The row and column indices in the matrix are denoted by $i$ and $j$, respectively.
  \item [2)]
Calculate the asymmetric distortion $\rho_{}^+$, $\rho_{}^-$. Then, extract a cover sequence and the modification distances $d^+$, $d^-$ from DCT blocks using the generalized dither modulation algorithm.

  \item [3)]
Calculate the modifying costs $\xi^+$ and $\xi^-$
as shown in \eqref{con:dist}, where $\zeta_{ij}$ represents the distortion of the $ij$th dequantized DCT coefficient.
\begin{equation}
\label{con:dist}
\begin{aligned}\zeta_{ij}^+&=\frac{\rho_{ij}^+}{q_{ij}},\quad \xi_{ij}^+=\zeta_{ij}^+\times d_{ij}^+\\\zeta_{ij}^-&=\frac{\rho_{ij}^-}{q_{ij}},\quad \xi_{ij}^-=\zeta_{ij}^-\times d_{ij}^-\end{aligned}\end{equation}
  \item [4)]
The message $m$ is encoded using RS \citep{macwilliams1977theory} codes to produce an encoded message $m'$. Following this, the encoded message $m'$ is embedded into the cover sequence using ternary STC \citep{filler2011minimizing} to create a stego sequence. The stego sequence and modification distances $d^+$, $d^-$ are then employed to generate a stego image by quantizing the DCT coefficients of the cover image.

\end{itemize}

\begin{figure*}[t]
    \centering
    \includegraphics[width=1.0\linewidth]{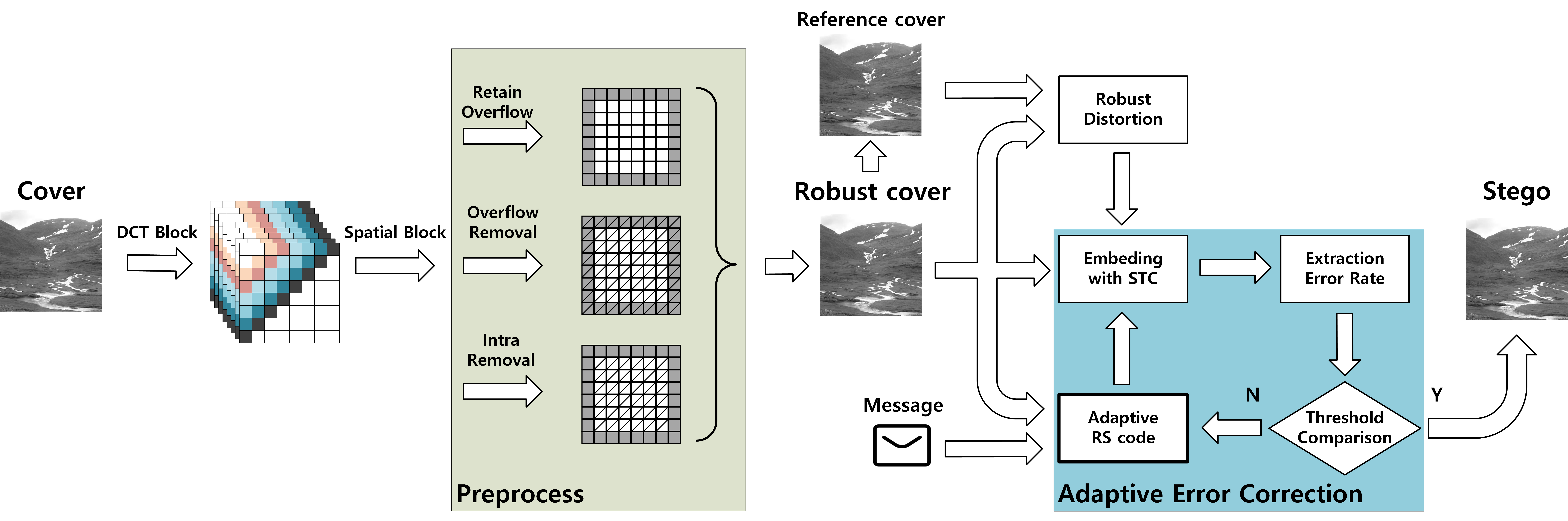}
    \caption{The embedding framework of the proposed method.}
    \label{fig:framework}
\end{figure*}  

To obtain the quantized DCT coefficients from the stego image, the receiver must employ the identical quantization table utilized during the embedding process. The encrypted data can then be extracted through STC \citep{filler2011minimizing} decoding. Finally, the secret information can be retrieved via RS \citep{macwilliams1977theory} decoding.

\subsection{ROAST}
Zeng et al. \citep{10093140} argue that spatial truncation, a lossy operation in JPEG recompression, significantly affects the robustness of current upward robust steganographic techniques. Therefore, within the framework of the GMAS \citep{yu2020robust} algorithm, the ROAST method has been proposed.

Two approaches have been proposed in ROAST. The first method, called ROAST-OS, eliminates overflow by globally scaling the overflowed 8 × 8 spatial blocks, thereby ensuring robustness. However, this method may impact the anti-steganalysis performance due to significant modifications.

Another method, called ROAST-ST, is a targeted truncation approach that adjusts overflow locations and handles intensity in a parameterized manner, resulting in improved the anti-steganalysis performance. The specific operation is shown in formula \eqref{con:st}, where $\hat{s}_{i,j}$ represents the corresponding pixel value obtained after spatial overflow preprocessing. Since it is difficult to completely eliminate overflow with a single execution of the operation in formula \eqref{con:st}, this method will also perform an additional overflow removal operation to obtain a reference cover.
\begin{equation}
\label{con:st}
\hat{s}_{i,j}=\begin{cases}127-\textit{T1}&\quad s_{i,j}>127\\-128+\textit{T1}&\quad s_{i,j}<-128\end{cases}\end{equation}

This method presents a new robust asymmetric distortion technique that integrates the pre-overflow removal operation into the information embedding process. It handles positions where the reference cover coefficients exceed or fall below those of the robust cover with lower costs of '+1' and '-1', respectively.

In summary, the ROAST method exhibits high robustness by effectively mitigating spatial overflow. Additionally, by enhancing the stability of quantized DCT coefficients, the entire image can be regarded as a robust region, significantly increasing the embedding capacity.



\section{Proposed method}
\label{section:Proposed method}
This section introduces the overall framework, analyzes the characteristics of spatial block overflow, and provides detailed explanations of the preprocessing method and adaptive error correction technique.

\begin{figure}[b]
    \centering
    \begin{subfigure}[b]{0.45\linewidth}
        \centering
        \includegraphics[width=\linewidth]{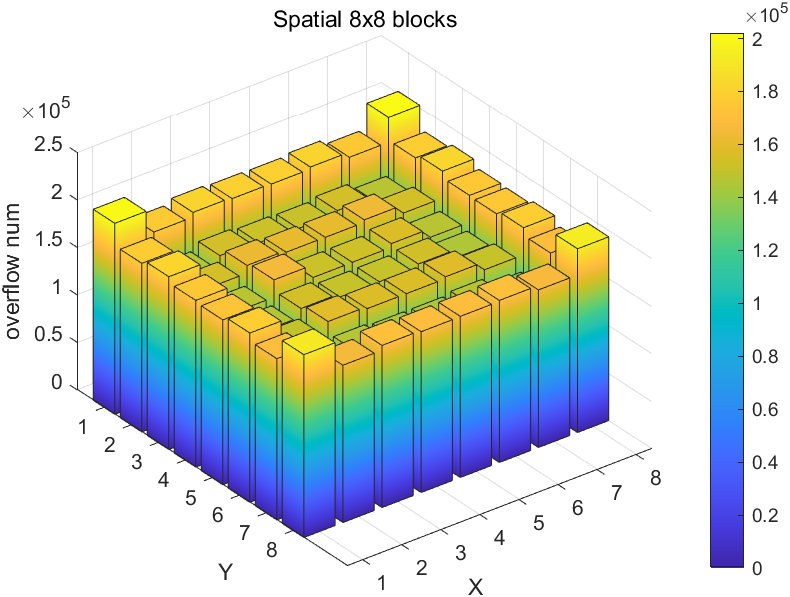}
        \caption{}
        \label{fig:overflow}
    \end{subfigure}
    \vspace{10pt}
    \quad
    \begin{subfigure}[b]{0.40\linewidth}
        \centering
        \includegraphics[width=\linewidth]{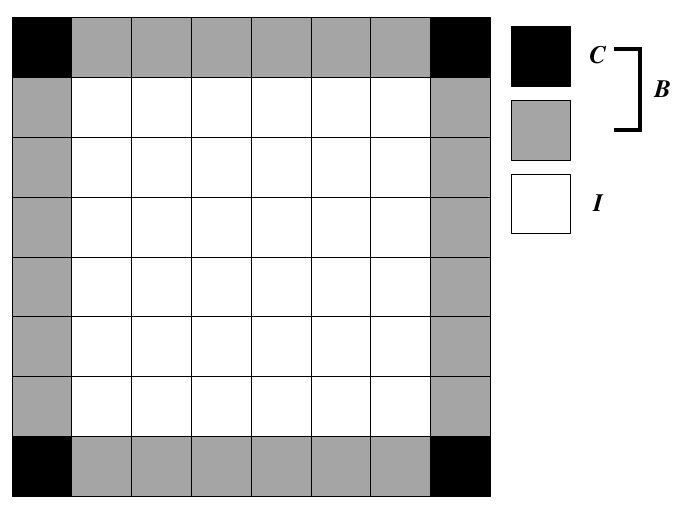}
        \caption{}
        \label{fig:dct}
    \end{subfigure}
    \caption{Introduction of the 8 × 8 spatial block. (a) The overflow situation of spatial 8 × 8 blocks; (b) Description of the different positions in the spatial 8 × 8 block.}
    \label{fig:introduction8*8block}
\end{figure}

\subsection{Framework of the Proposed Scheme}

Fig. \ref{fig:framework} illustrates the embedding framework of the proposed method. The \textit{Cover} undergoes an IDCT transformation to generate its spatial domain image. Preprocessing operations are then applied to each 8 × 8 block of the spatial domain to obtain the \textit{Robust cover}. Following the approach proposed in ROAST-ST \citep{10093140}, an overflow removal operation is performed to create the \textit{Reference cover}. Calculate the corresponding asymmetric distortion, handling positions where the \textit{Reference cover} exceeds or falls below the \textit{Robust cover} with smaller costs of '+1' and '-1', respectively. Finally, adaptive error correction coding is used, and ternary STC \citep{filler2011minimizing} embedding is performed to obtain the stego image, \textit{Stego}. The secret information extraction process is consistent with Section \ref{section:GMAS}.

\subsection{Minimizing Boundary Modifications in Spatial Block Preprocessing}

\begin{figure*}[ht]
    \centering
    \includegraphics[width=0.74\linewidth]{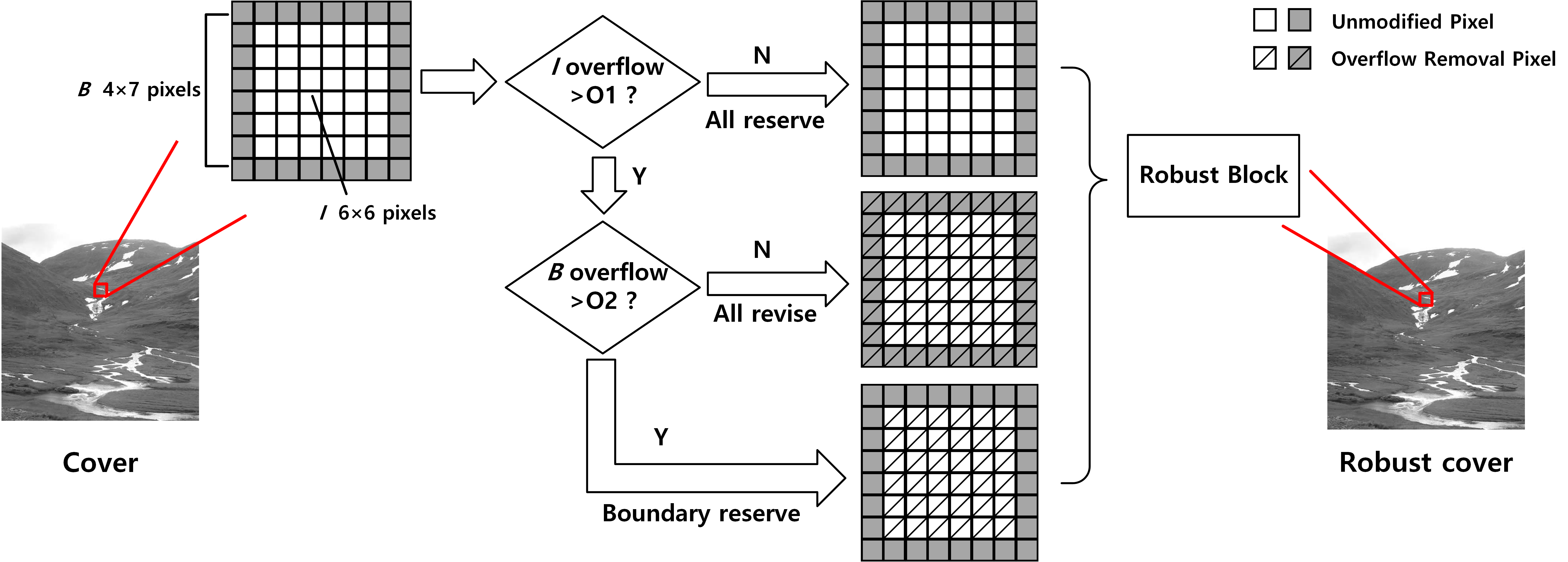}
    \captionsetup{justification=centering}
    \caption{The steps of preprocessing.}
    \label{fig:preprocessing}
\end{figure*}

\begin{figure}[b]
    \centering
    \begin{subfigure}[b]{0.45\linewidth}
        \centering
        \includegraphics[width=\linewidth]{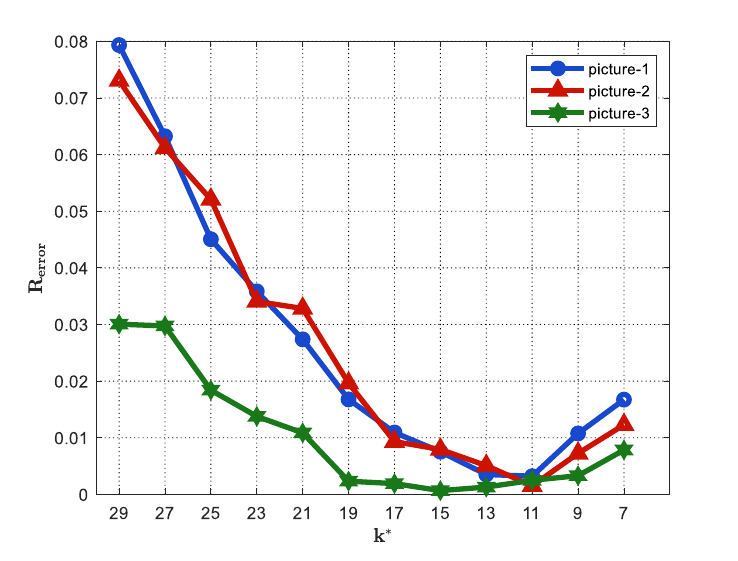}
        \caption{}
        \label{error_rs}
    \end{subfigure}
    \hspace{0.01\linewidth} 
    \begin{subfigure}[b]{0.43\linewidth}
        \centering
        \includegraphics[width=\linewidth]{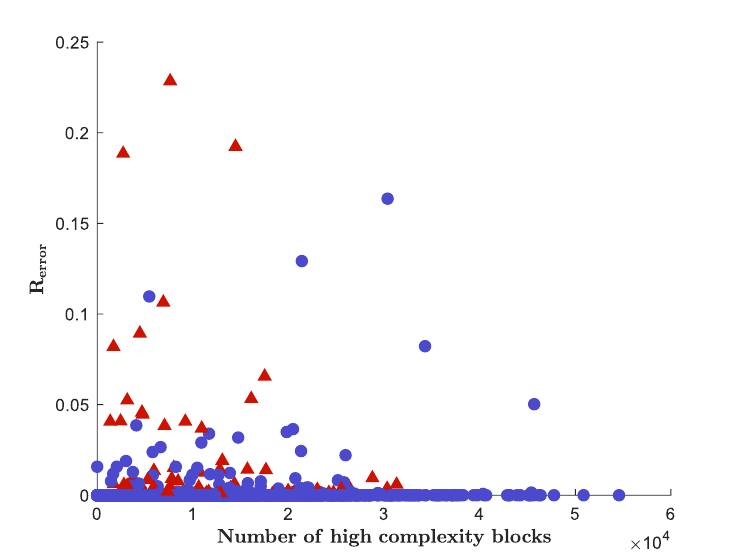}
        \caption{}
        \label{complexity}
    \end{subfigure}
    \caption{Abnormal circumstances occur during adaptive error correction. (a) Different image error rates under various error correction codes ($n^{\ast}$ = 31); (b) Exceptional conditions of different complexity.}
    \label{fig:rs}
\end{figure}


Existing preprocessing methods may significantly alter image pixels while removing overflow, thereby affecting the anti-steganalysis performance of steganography. To mitigate this impact, a statistical analysis was conducted on 10,000 images from the BOSSbase v1.01 \citep{bas2011break}, each compressed with a QF of 65, to illustrate the overflow situations of spatial 8 × 8 blocks, as depicted in Fig. \ref{fig:introduction8*8block}\subref{fig:overflow}.

The data indicates that overflow is most prominent at the boundaries, suggesting that removing all overflow from the 8 × 8 blocks would require significant modifications to the boundary regions. Wang et al. \citep{wang2020non} argue that the empirical security of spatial block boundaries is lower than that of the block interiors, implying that modifications to spatial block boundaries are more easily detectable than those made to the interiors. Furthermore, as the boundaries of spatial blocks most directly reflect the correlation between blocks, which reflects the similarity of local features in the image, minimizing alterations to these boundaries during preprocessing helps maintain image quality and reduce the risk of steganalysis detection. However, given that overflow situations are more apparent at the block boundaries, neglecting overflow removal operations at these boundaries would affect the stability of the quantized DCT coefficients, thus compromising the robustness. Therefore, effectively handling overflow at block boundaries is a crucial issue for improving the performance of the steganographic technique.

In response to the aforementioned issue, we have proposed a preprocessing method that minimizes modifications to the block boundaries by removing overflow based on the actual overflow situation of the blocks. For ease of describing the preprocessing method, Fig. \ref{fig:introduction8*8block}\subref{fig:dct} illustrates the configuration of the 8 × 8 spatial blocks in the image. The corners of these blocks are marked in black and denoted as \textbf{\textit{C}}. The boundaries of the spatial blocks consist of 28 pixels along the four edges (including the corners) and are represented as \textbf{\textit{B}}. The internal 6 × 6 blocks, containing 36 pixels, are denoted as \textbf{\textit{I}}.

Fig. \ref{fig:preprocessing} illustrates our preprocessing algorithm flow. The areas marked with diagonal lines indicate that overflow removal operations have been performed in those regions. For an image, overflow checks are initially conducted on the \textit{\textbf{I}} parts of each spatial 8 × 8 block. A decision is made on whether to perform overflow removal on the \textit{\textbf{I}} blocks based on their overflow situation. The parameter \textit{O1} measures the overflow of the \textbf{\textit{I}} part in each block; if the number of overflow pixels exceeds \textit{O1}, the \textbf{\textit{I}} part pixels are processed as described in formula \eqref{con:st}. If an overflow removal operation is performed on part \textbf{\textit{I}}, the \textit{\textbf{B}} parts of the 8 × 8 blocks to which the \textit{\textbf{I}} parts belong are checked for overflow. Based on the overflow situation, a decision is made whether to perform overflow removal on the \textit{\textbf{B}} parts to complete the spatial block preprocessing. The parameter \textit{O2} is used to measure the overflow of part \textbf{\textit{B}} in each block, and if the number of overflowed pixels is less than \textit{O2}, the pixels in part \textbf{\textit{B}} undergo the same overflow removal operations. Compared to ROAST-ST \citep{10093140}, we reduce modifications to block boundaries to enhance anti-steganalysis performance while maintaining robustness.

\subsection{Adaptive Error Correction }

For images with strong robustness, excessively long error correction codes can diminish the anti-steganalysis performance by introducing redundancy in error correction capability. Conversely, for images with weak robustness, overly short error correction codes may not provide significant error correction effects. $(n^{\ast},k^{\ast})$RS \citep{macwilliams1977theory} codes are commonly used in steganography, where  $n^{\ast}$ and $k^{\ast}$ represent the code length and message length, respectively. In experiments, the code length $n^{\ast}$ is generally set to 31. The greater the ratio of $k^{\ast}$ to $n^{\ast}$, the stronger the error correction ability of RS codes. GMAS \citep{yu2020robust} employs fixed RS error correction codes, while Adaptive-GMAS \citep{duan2023robust} adaptively utilizes 12 different RS error correction capabilities. The adaptive error correction coding approach effectively addresses the aforementioned issues by ensuring that different images fully leverage error correction codes.

However, in practical applications, different images undergo varying degrees of loss and exhibit different levels of robustness following JPEG compression. For instance, employing a payload of 0.2, we conducted an evaluation on 10,000 images sourced from the BOSSbase v1.01 \citep{bas2011break} to assess their robustness. This evaluation involved the utilization of various RS \citep{macwilliams1977theory} error correction codes in conjunction with our preprocessing method, alongside the ROAST-ST \citep{10093140}. The relative payload is defined as $n_{\mathfrak{m}}/n _ {\text{nzac}}$, where $n _ {\text{nzac}}$ represents the number of nonzero AC DCT coefficients in the original cover image, rather than in the robust cover image after de-overflow processing and  $n_{\mathfrak{m}}$ represents the total number of message bits. The robustness of an image is commonly represented by the extraction error rate. A higher extraction error rate indicates poorer robustness. The extraction error rate is calculated as $\begin{aligned}R_{\text{error}} = n _ {\text{error}}/ n _ {\text{m}}\end{aligned}$, where $n _ {\text{error}}$ represents the number of incorrect  message bits. 

The experimental findings indicate a minor elevation in the error rates of select images as the length of the error correction codes extends. Three images demonstrating this phenomenon were randomly selected, as depicted in Fig. \ref{fig:rs}\subref{error_rs}. To further investigate this occurrence, a subset of 2000 images was randomly selected from a pool of 10,000 images, and their texture complexity was evaluated using the method proposed by Zhang et al. \citep{zhang2019multiple}, as depicted in Fig. \ref{fig:rs}\subref{complexity}. The horizontal axis indicates the count of high-complexity blocks per image when the image is divided into multiple 2 × 2 blocks, while the vertical axis represents the error rate of the image. Images with error rates that do not decline with increasing error correction code length are denoted by triangles, whereas those exhibiting the opposite trend are represented by circles. We observed that the images showing this phenomenon mainly consist of images with smooth textures. This is because the algorithm extends the embedding domain to the entire DCT domain, prioritizing embedding in low-frequency regions, which typically have lower distortion costs during embedding. However, smooth regions are less suitable for robust information embedding. Therefore, as the length of the error correction code increases, more bits are embedded in these regions, thereby leading to an elevation in the error rate.

\begin{algorithm}[t]
\small
\renewcommand\arraystretch{1.4}
\scriptsize
\caption{Adaptive RS Code}\label{alg:alg1}
\KwIn{cover image $\boldsymbol{cover}$, message $\boldsymbol{m}$, \\ 
      error rate $\boldsymbol{threshold}$, channel quality $\boldsymbol{Q_{channel}}$}
\KwOut{optimal RS code $\boldsymbol{RS(31,best\_k^*)}$, stego image $\boldsymbol{stego}$}
\textit{Initial RS code}
$\boldsymbol{k^*} \leftarrow 29$ \; \textit{Initial error rate}
$\boldsymbol{R_{error}} \leftarrow 1$ \; 
$\boldsymbol{best\_R_{error}} \leftarrow 1$ \; 
\While{$\boldsymbol{R_{error}} > \boldsymbol{threshold}$ \textbf{or} $\boldsymbol{k^*} \geq 7$}{
    $\boldsymbol{msg} \leftarrow \text{RS\_Encode}(\boldsymbol{m}, 31, \boldsymbol{k^*})$\;
    $\boldsymbol{stego} \leftarrow \text{STC3Embed}(\boldsymbol{S}, \boldsymbol{msg})$\;
    $\boldsymbol{stego'} \leftarrow \text{Recompression}(\boldsymbol{stego}, \boldsymbol{Q_{channel}})$\;
    $\boldsymbol{m'} \leftarrow \text{RS\_Decode}(\text{STC3Extract}(\boldsymbol{stego'}))$\;
    $\boldsymbol{R_{error}} \leftarrow \text{Calculate\_Error\_Rate}(\boldsymbol{m}, \boldsymbol{m'})$\;

    \If{$\boldsymbol{R_{error}} < \boldsymbol{best\_R_{error}}$}{
        $\boldsymbol{best\_k^*} \leftarrow \boldsymbol{k^*}$\;
    }

    \If{$\boldsymbol{R_{error}} > \boldsymbol{threshold}$}{
        $\boldsymbol{k^*} \leftarrow \boldsymbol{k^*} - 2$\;
    }
}
\end{algorithm}

In response to the aforementioned scenario, we have refined the existing adaptive error correction algorithm. Algorithm \ref{alg:alg1} presents the pseudocode for our improved adaptive error correction algorithm. Essentially, each iteration compares the error rate obtained with the previous best error rate using adaptive error correction. If the current error rate is lower than the best error rate, the best error rate is updated to the error rate obtained in this iteration. This process continues until $k^{\ast}$ equals 7 or the error rate reaches a preset threshold. This method effectively addresses the challenge of increasing error correction code lengths without compromising error rates. Additionally, while maintaining robustness, the prospective adoption of shorter error correction codes is anticipated to enhance anti-steganalysis performance.

\section{Experiment}
\label{section:Experiment}

\subsection{Setups}
The proposed method was evaluated using the BOSSbase v1.01 dataset \citep{bas2011break}, which consists of 10,000 grayscale images with a resolution of 512 × 512 pixels, consistent with the dataset used by ROAST \citep{10093140}. This dataset offers a comprehensive assessment of the method's performance. To facilitate evaluation, we randomly generated 2,000 serial numbers from the range of 10,000 using the '\textbf{randperm}' function in MATLAB with a random seed of 25. Corresponding to these serial numbers, 2,000 images were selected and compressed with JPEG at a QF of 65 to create cover images. To simulate various channel compression conditions, the $Q_{channel}$ were set to 85 and 95. 

\begin{table}
\centering
\renewcommand\arraystretch{1.5}
\scriptsize
\caption{Settings}
\label{tab:setup}
\begin{tabular}{l  l}
\hline
   Image source & BOSSbase v1.01 \citep{bas2011break}  \\
  Image size    & $512 \times 512$  \\
  Number of images  & $2000$    \\
  Random seed   & 25  \\
  Secret message & Random binary sequence\\
  Choice of parameter $ \textit{T1},\mu$   & consistent with ROAST-ST  \\
  STC secure parameter $h$ & 10\\
  Choice of robustness thershold  & 0.0001  \\
\hline

\end{tabular}
\end{table}

\begin{figure*}[b]
    \centering
    \begin{subfigure}[b]{0.23\linewidth}
        \centering
        \includegraphics[width=\linewidth]{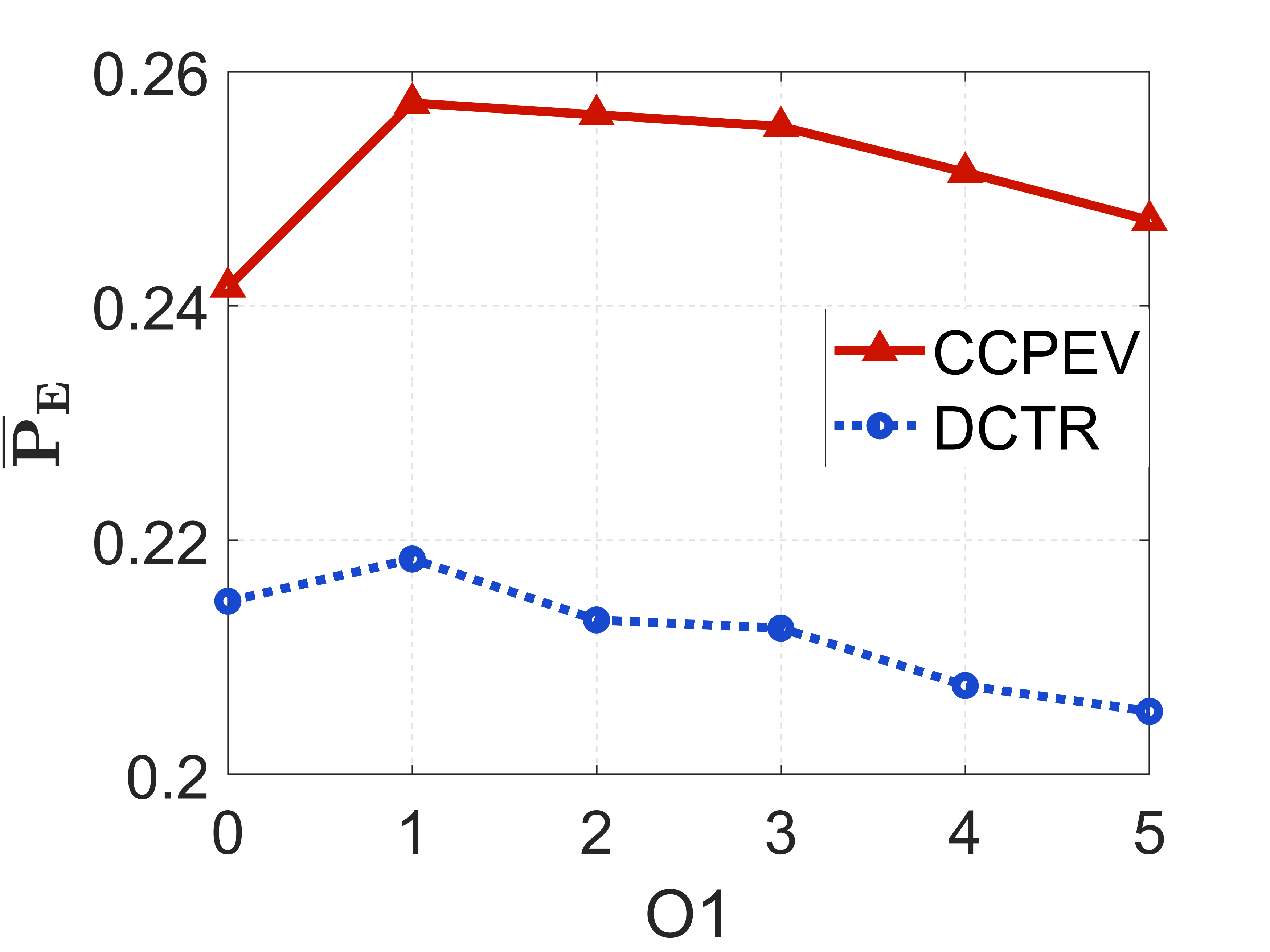}
        \caption{Security of O1}
        \label{o1_security}
    \end{subfigure}
    \hspace{0.01\linewidth} 
    \begin{subfigure}[b]{0.23\linewidth}
        \centering
        \includegraphics[width=\linewidth]{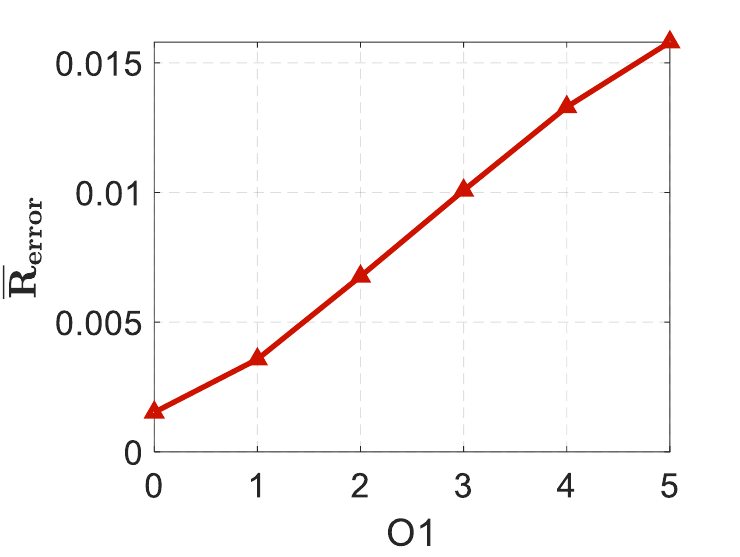}
        \caption{Robustness of O1}
        \label{o1_robustness}
    \end{subfigure}
    \quad
    \begin{subfigure}[b]{0.23\linewidth}
        \centering
        \includegraphics[width=\linewidth]{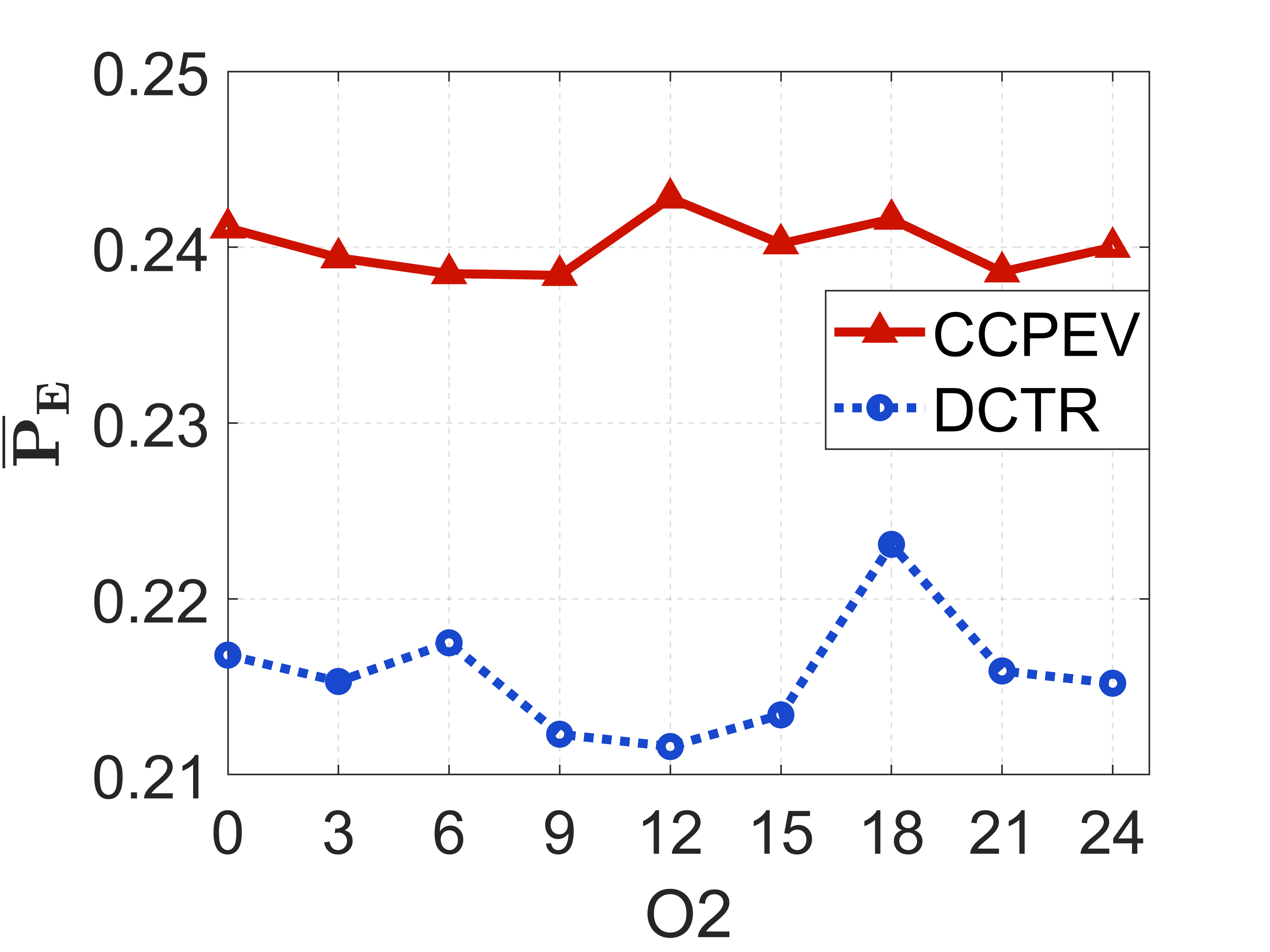}
        \caption{Security of O2}
        \label{o2_security}
    \end{subfigure}
    \hspace{0.01\linewidth} 
    \begin{subfigure}[b]{0.23\linewidth}
        \centering
        \includegraphics[width=\linewidth]{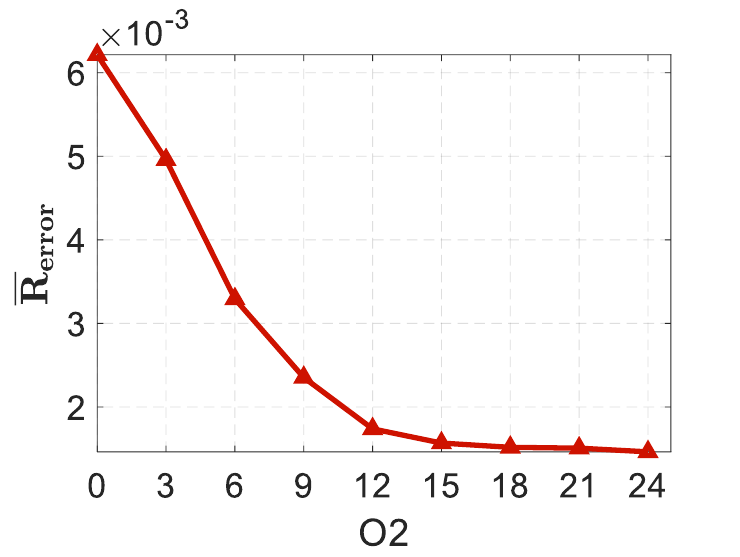}
        \caption{Robustness of O2}
        \label{o2_robustness}
    \end{subfigure}
    \caption{Different performance affected by O1 and O2 at $Q_{channel}$ = 85.}
    \label{fig:parameter}
\end{figure*}

Previous steganographic techniques have exhibited limited performance, typically setting the range of robust adaptive steganography from 0.05 to 0.15 bits per nonzero AC coefficient (\textbf{bpnzac}). By eliminating the need for robust domain selection after preprocessing, our method achieves a high embedding rate, allowing experiments in this paper to encompass payloads ranging from 0.1 to 0.5 bpnzac. Additionally, we select two effective feature sets (DCTR \citep{holub2014low}, CCPEV \citep{kodovsky2009calibration}) and SRNet \citep{8470101} to test the anti-steganalysis performance. Steganalysis is conducted using FLD ensemble \citep{kodovsky2011ensemble} classifiers, with separate classifiers trained for each algorithm and payload combination. By default, the ensemble minimizes the total classification error probability under equal priors, where $P_\mathrm{E}=\min_{P_\mathrm{FA}}\frac{1}{2}(P_\mathrm{FA}+P_\mathrm{MD})$, with $P_\mathrm{FA}$ and $P_\mathrm{MD}$ denoting the false-alarm probability and the missed-detection probability, respectively. Training SRNet involves processing cover and stego images for 300 epochs with an initial learning rate of r1 = 0.001, followed by an additional 100 epochs with a learning rate of r2 = 0.0001. For each embedding algorithm and payload, a separate classifier and deep network are trained. The overall security is determined by the average error rate $\bar{P}_{\mathrm{E}}$, with a higher $\bar{P}_{\mathrm{E}}$ indicating stronger security. We select the STC secure parameter \textit{h} = 10 to achieve efficient steganography with minimal distortion and set the robustness threshold to 0.0001, consistent with that of Adaptive-GMAS \citep{duan2023robust}, to ensure accurate information extraction. The other parameters of the proposed method remain consistent with ROAST-ST \citep{10093140}, such as the \textit{T1} = 8, $\mu$ = 0.5, etc. The experimental setup is summarised in Table \ref{tab:setup}, providing the key parameters and configurations used in our experiments. To ensure the reproducibility and stability of the experimental results, we conducted multiple independent repeat experiments.


\subsection{Exploring the Performance of Preprocessing}

In the preprocessing experiments, we assessed the overflow of part \textbf{\textit{I}} versus part \textbf{\textit{B}} using two parameters: \textit{O1} and \textit{O2}, respectively. In this section, we examine the impact of these parameters on the algorithm's robustness and anti-steganalysis performance to facilitate subsequent performance testing.

\begin{itemize}

\begin{figure*}[b]
\centering
\subfloat[$Q_{channel} = 85$ CCPEV]{
\includegraphics[width=0.35\linewidth]{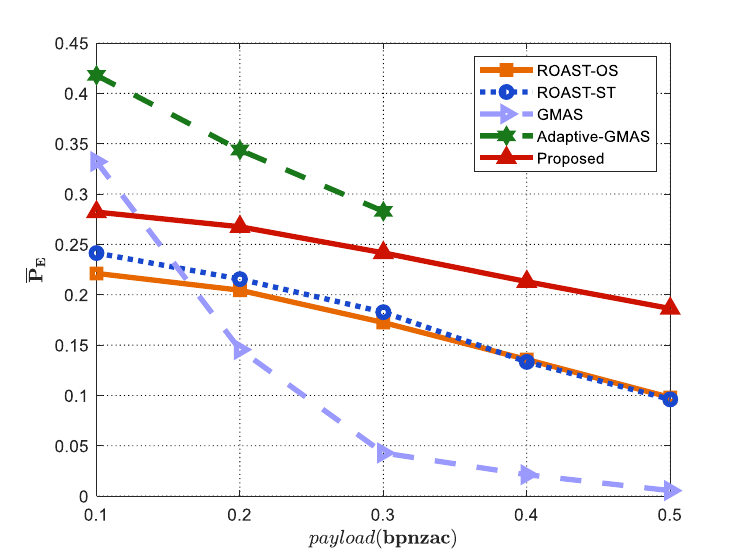} 
}
\subfloat[$Q_{channel} = 85$ DCTR]{
\includegraphics[width=0.35\linewidth]{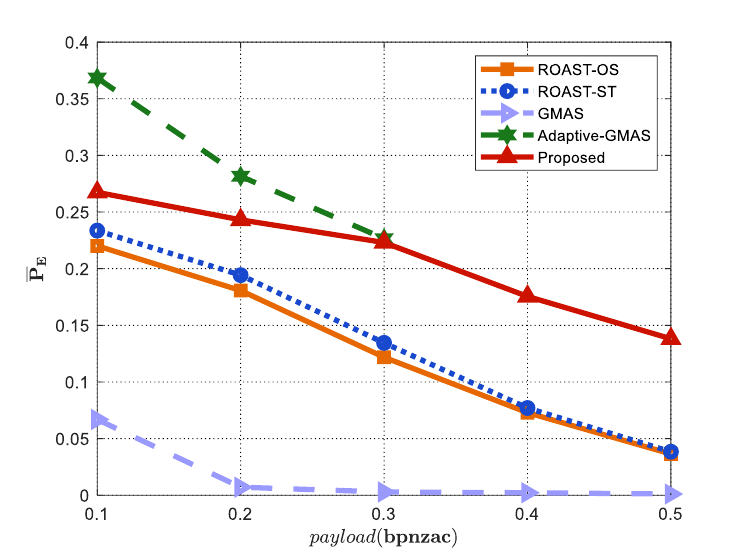} 
}
\quad
\subfloat[$Q_{channel} = 95$ CCPEV]{
\includegraphics[width=0.35\linewidth]{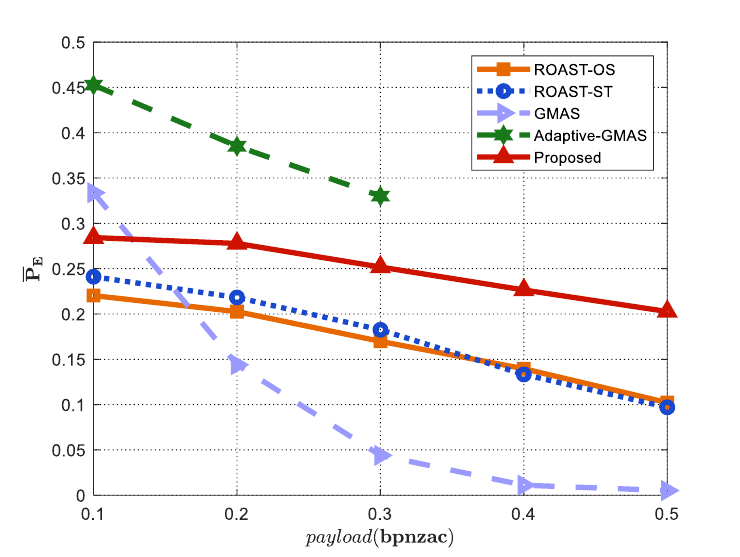}
}
\subfloat[$Q_{channel} = 95$ DCTR]{
\includegraphics[width=0.35\linewidth]{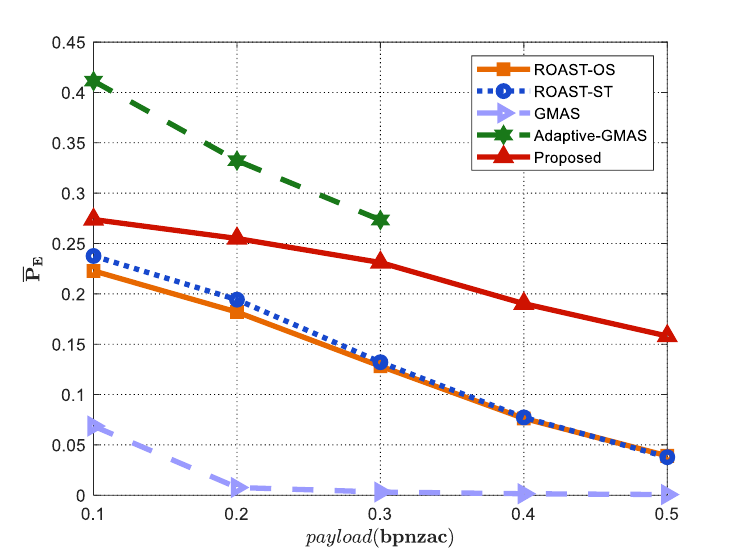}
}
\caption{Comparison of anti-steganalysis performance at $Q_{channel}$ = 85 and $Q_{channel}$ = 95.}
\label{fig:security}
\end{figure*}

  \item [1)]
\textit{Effect of Parameter O1:} The parameter \textit{O1} is used to evaluate the overflow condition of part \textbf{\textit{I}}. The impact on the algorithm's robustness and security is depicted in Fig. \ref{fig:parameter}\subref{o1_security} and \ref{fig:parameter}\subref{o1_robustness}. It is evident that \textit{O1} significantly influences the algorithm's robustness, with the error rate increasing as \textit{O1} values escalate. Considering all factors, since \textit{O1} has a minimal effect on anti-steganalysis performance, we set it to 0 in subsequent experiments.

  \item [2)]
\textit{Effect of Parameter O2:} The \textit{O2} parameter evaluates overflow occurrences at block boundaries. When the number of overflow pixels at the block boundary exceeds \textit{O2}, conducting de-overflow processing on the boundary could significantly alter the block, disrupting the correlation between neighboring blocks. Consequently, in such scenarios, de-overflow processing is omitted at this boundary. Fig. \ref{fig:parameter}\subref{o2_security} and \ref{fig:parameter}\subref{o2_robustness} demonstrate that parameter \textit{O2} has a minor effect on steganalysis performance but a significant impact on the robustness of the algorithm. Specifically, as \textit{O2} increases from 0 to 18, the average error rate decreases significantly. However, when \textit{O2} is greater than or equal to 18, the algorithm's robustness remains almost unchanged. Furthermore, at this value, the algorithm exhibits high anti-steganalysis performance. Therefore, we have selected \textit{O2} = 18 for subsequent experiments.


\end{itemize}


\begin{table}[t]
\centering 
\scriptsize
\caption{Comparison of image quality} 
\label{tab:performance}
\setlength{\tabcolsep}{2pt} 
\renewcommand\arraystretch{1.5}
\begin{tabular}{ >{\centering\arraybackslash}m{1.5cm}  >{\centering\arraybackslash}m{1cm}  >{\centering\arraybackslash}m{1cm}  >{\centering\arraybackslash}m{1cm}  >{\centering\arraybackslash}m{1cm}  >{\centering\arraybackslash}m{1cm}  >{\centering\arraybackslash}m{1cm} }
\toprule
                      & \textit{\textbf{payload}} & \textbf{\textit{ROAST-OS \citep{10093140}}} & \textbf{\textit{ROAST-ST \citep{10093140}}} & \textbf{\textit{GMAS \citep{yu2020robust}}} & \textbf{\textit{Adaptive-GMAS \citep{duan2023robust}}} & \textbf{\textit{Proposed}} \\ \midrule
\multirow{2}{*}{\textbf{PSNR}} & 0.2     & 42.079                     & 42.815                     & 40.222                 & \textbf{46.262}                         & \textbf{44.471}                       \\ 
                      & 0.4     & 40.189                     & 40.702                     & 36.234                 & \textbackslash{}               & \textbf{42.330}                     \\ \midrule
\multirow{2}{*}{\textbf{SSIM}} & 0.2     & 0.9882                     & 0.9886                     & 0.9843                 & \textbf{0.9935}                         & \textbf{0.9915}                     \\ 
                      & 0.4     & 0.9827                     & 0.9832                     & 0.9539                 & \textbackslash{}               & \textbf{0.9868}                     \\ \bottomrule
\end{tabular}
\end{table}

\subsection{Comparison of Security}

In image steganographic techniques, security is typically divided into imperceptibility and anti-steganalysis performance. 


Image quality is a critical metric for evaluating the imperceptibility of steganographic techniques. TABLE \ref{tab:performance} compares the average Peak Signal-to-Noise Ratio (PSNR) and Structural Similarity (SSIM) values of different methods at embedding rates of 0.2 and 0.4. Since Adaptive-GMAS \citep{duan2023robust} may iterate to higher frequency embedding domains with longer error correction coding, the algorithm cannot be fully implemented on some images with embedding rates of 0.4 and 0.5. Consequently, except for Adaptive-GMAS, the stego images generated by the proposed method exhibit significantly higher image quality, indicating superior imperceptibility.



\begin{figure}[t]
    \centering
    \includegraphics[width=0.68\linewidth]{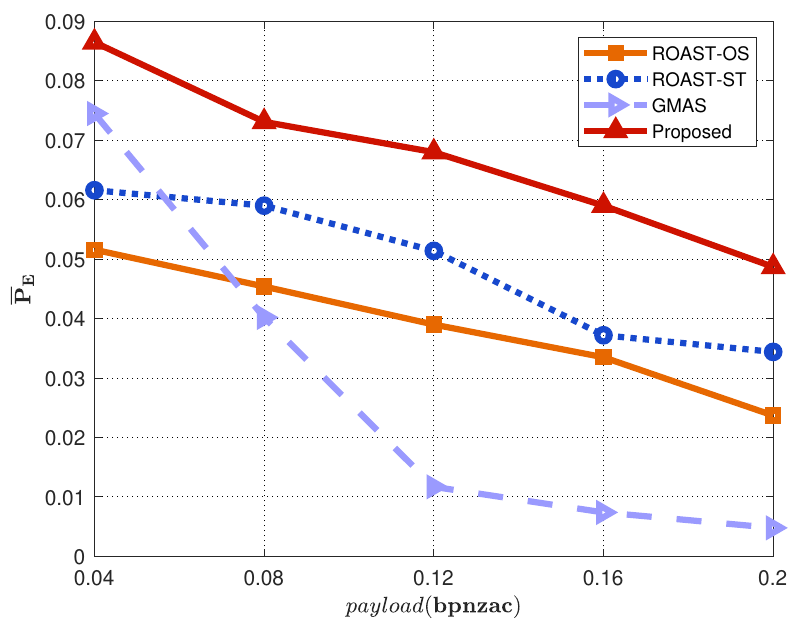}
    \captionsetup{justification=centering}
    \caption{Comparison of the security among various algorithms utilizing SRNet ($Q_{cover}$ = 65, $Q_{channel}$ = 85).}
    \label{fig:SRNet}
\end{figure}

\begin{figure*}[b]
    \centering
    \subfloat[$Q_{channel} = 85$]{
    \label{robust_85}
    \includegraphics[width=0.35\linewidth]{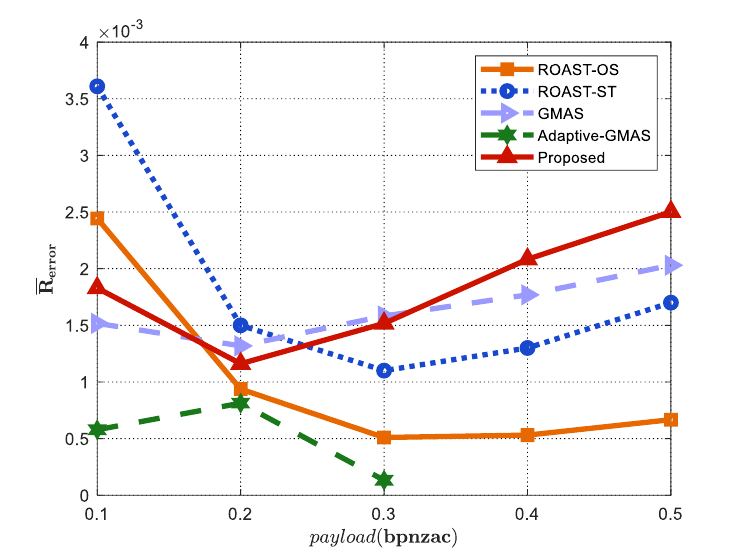}}
    \subfloat[$Q_{channel} = 95$]{
    \label{robust_95}
    \includegraphics[width=0.35\linewidth]{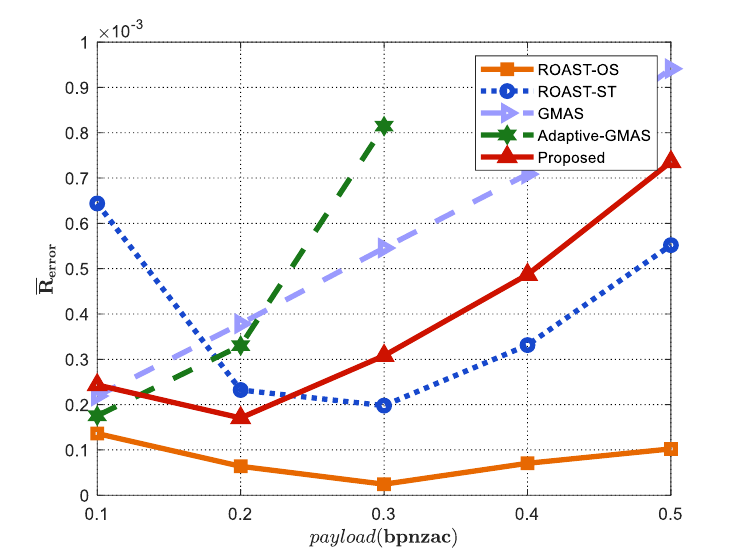}
    }
    \caption{Comparison of the robustness at (a) $Q_{channel}$ = 85 and (b) $Q_{channel}$ = 95.}
   \label{fig:robustness}
\end{figure*}

Fig. \ref{fig:security} illustrates the anti-steganalysis performance of various steganographic algorithms at $Q_{channel}$ = 85 and $Q_{channel}$ = 95. While Adaptive-GMAS \citep{duan2023robust} demonstrates relatively strong anti-steganalysis performance, its classification error rate decreases rapidly with increasing embedding rates, and the method suffers from a significantly constrained embedding capacity. The proposed method performs embedding of secret information across the entire DCT domain, resulting in a higher embedding rate. The proposed method demonstrates superior security compared to other methods, except for GMAS \citep{yu2020robust} at a payload of 0.1 using CCPEV \citep{kodovsky2009calibration}. 

When using the popular CNN-based steganalysis, we do not employ Adaptive-GMAS \citep{duan2023robust} with SRNet \citep{8470101} due to its limited embedding rate. Fig. \ref{fig:SRNet} compares the anti-steganalysis performance of different steganographic techniques at low embedding rates, demonstrating the superior performance of the proposed method. Compared to the ROAST-ST \citep{10093140} method, the proposed method achieves an over 2$\%$ increase in the average detection error rate at an embedding rate of 0.16.

Overall, the proposed method enhances security by minimizing modifications to block boundaries, thereby preserving more edges and image details. This leads to maintaining image quality and achieving better anti-steganalysis performance after preprocessing. In contrast, ROAST (ROAST-OS and ROAST-ST) \citep{10093140} may cause significant modifications to image pixels, thereby affecting anti-steganalysis performance. Like ROAST, GMAS employs fixed length RS(31,15) \citep{macwilliams1977theory} encoding for the secret message. However, the encoding efficiency is low, and the anti-steganalysis performance decreases rapidly with an increase in payload. In contrast, the proposed method utilizes adaptive RS coding to maximize the error correction capacity in different images, preventing the decrease in security due to redundant error correction codes.  


\subsection{Comparison of Robustness}

This subsection compares the robustness of the proposed method with other steganographic techniques.

For $Q_{channel}$ = 85, as shown in Fig. \ref{fig:robustness}\subref{robust_85}, the proposed method demonstrates greater robustness than the ROAST-ST \citep{10093140} method at low payload. This is primarily because of the adaptive error correction, which fully utilises the capabilities of the error-correcting code in various images to ensure robustness. Due to the overall overflow removal operation performed by ROAST-OS \citep{10093140} on the entire spatial block, it exhibits better robustness compared to ROAST-ST. As the payload increases to 0.3 or higher, the proposed method preprocesses by reducing modifications to block boundaries, which decreases the stability of the quantized DCT coefficients and results in slightly less robustness compared to the ROAST method. The robustness of the GMAS \citep{yu2020robust} method is similar to the proposed method. Adaptive-GMAS \citep{duan2023robust} exhibits high robustness at low embedding rates, but its embedding capacity is relatively small.

\begin{figure*}[t]
    \centering

    \subfloat[CCPEV]{
    \includegraphics[width=0.28\linewidth]{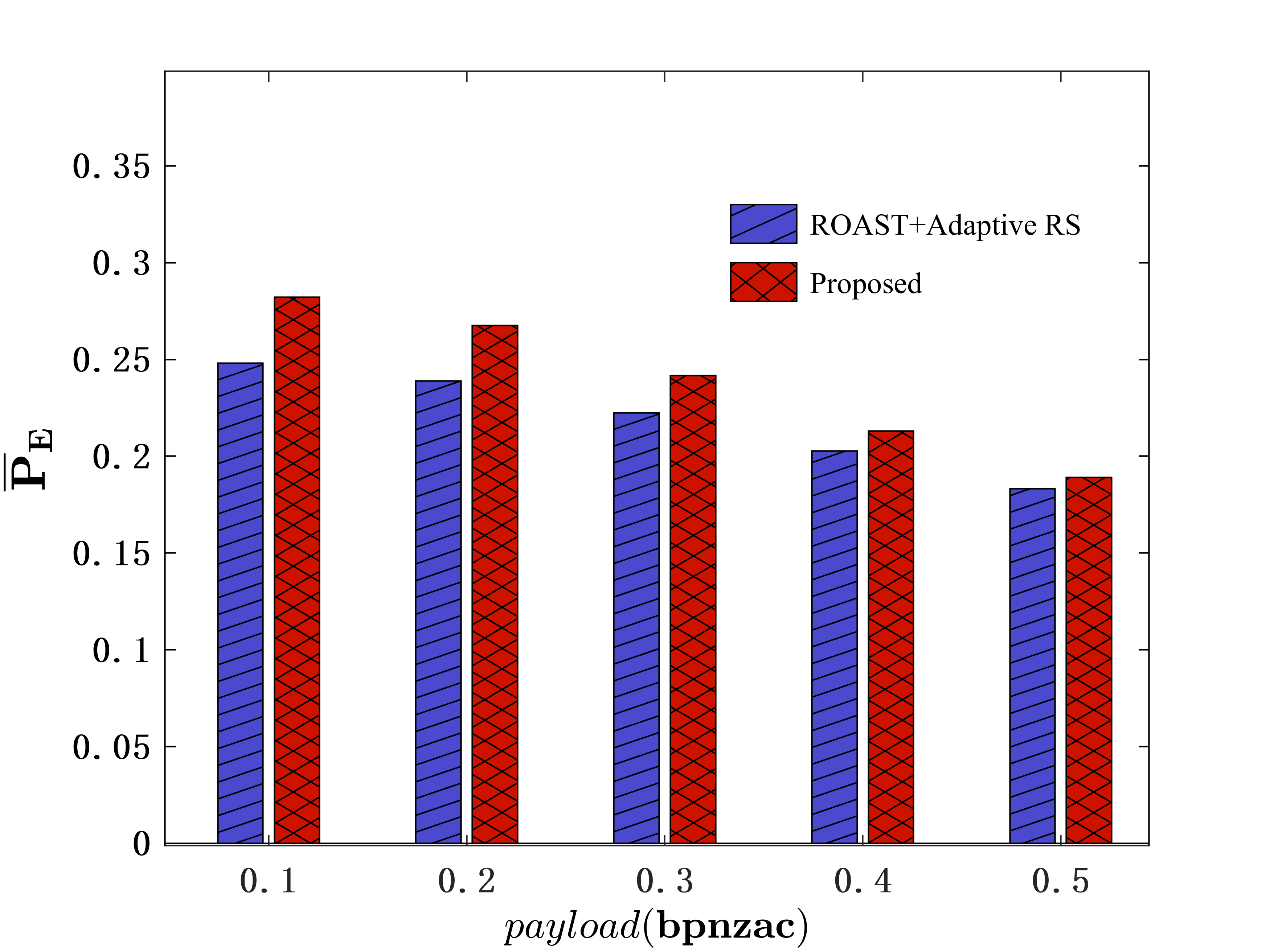}}
    \subfloat[DCTR]{
    \includegraphics[width=0.28\linewidth]{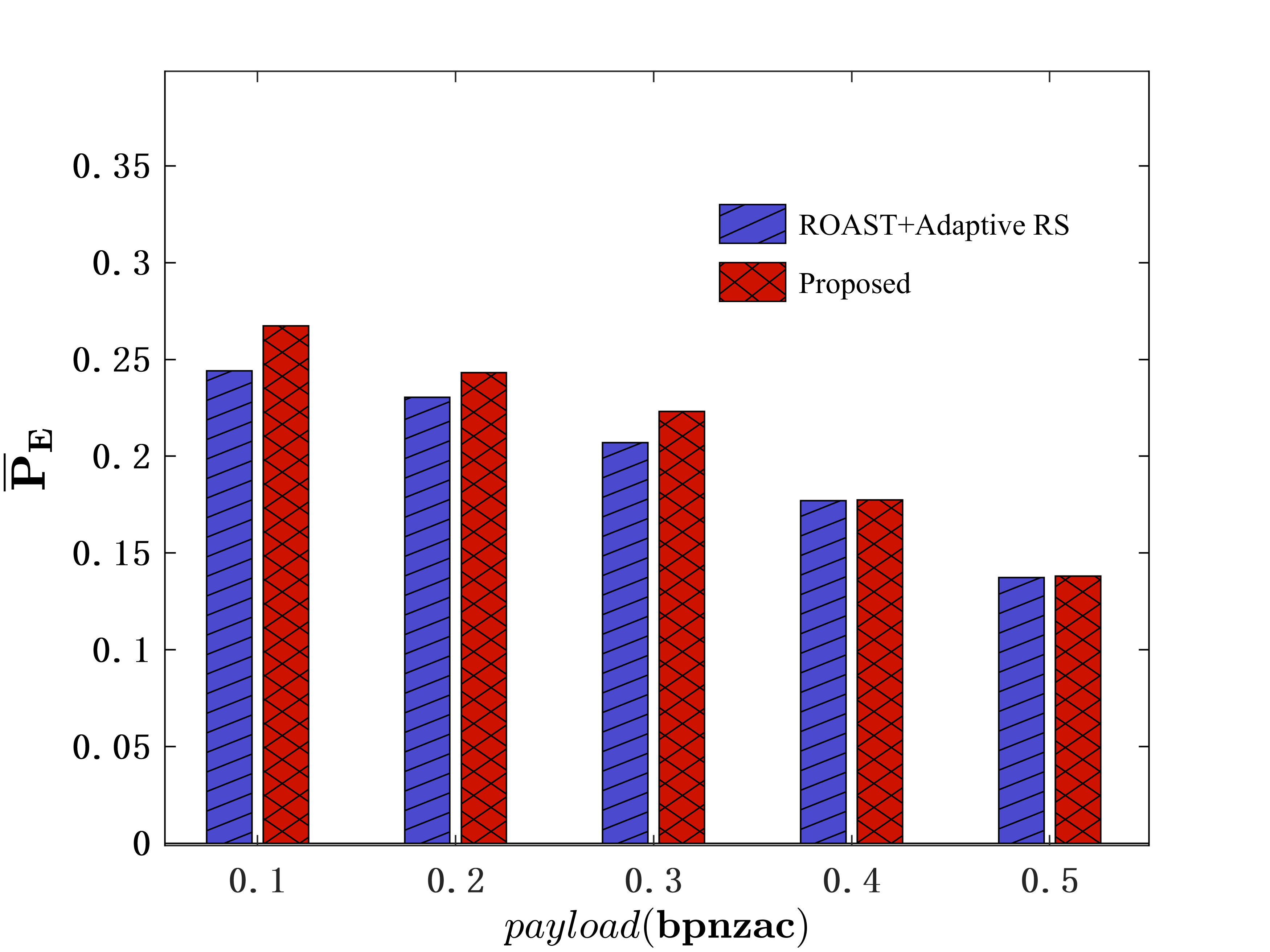}
    }
    \subfloat[Robustness]{
    \includegraphics[width=0.28\linewidth]{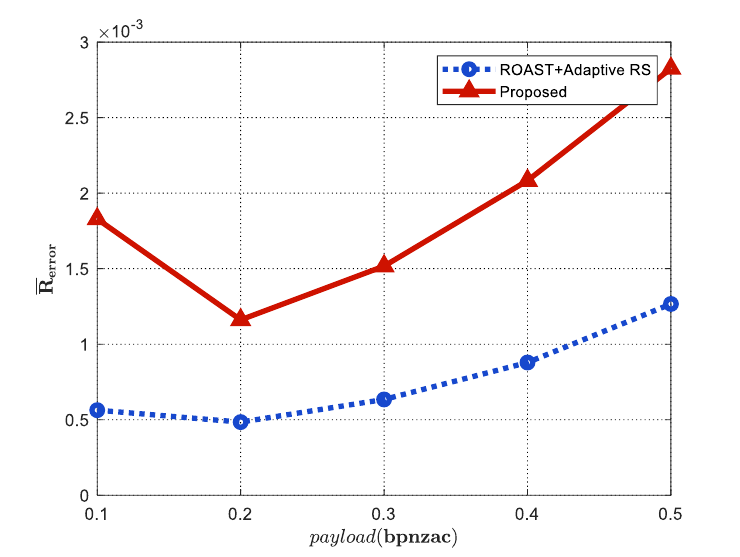}
    }
    \caption{Comparison of the performance about Preprocessing at $Q_{channel}$ = 85.}
    \label{fig:ab1}
\end{figure*}

\begin{figure*}[ht]
    \centering
    \subfloat[CCPEV]{
    \includegraphics[width=0.28\linewidth]{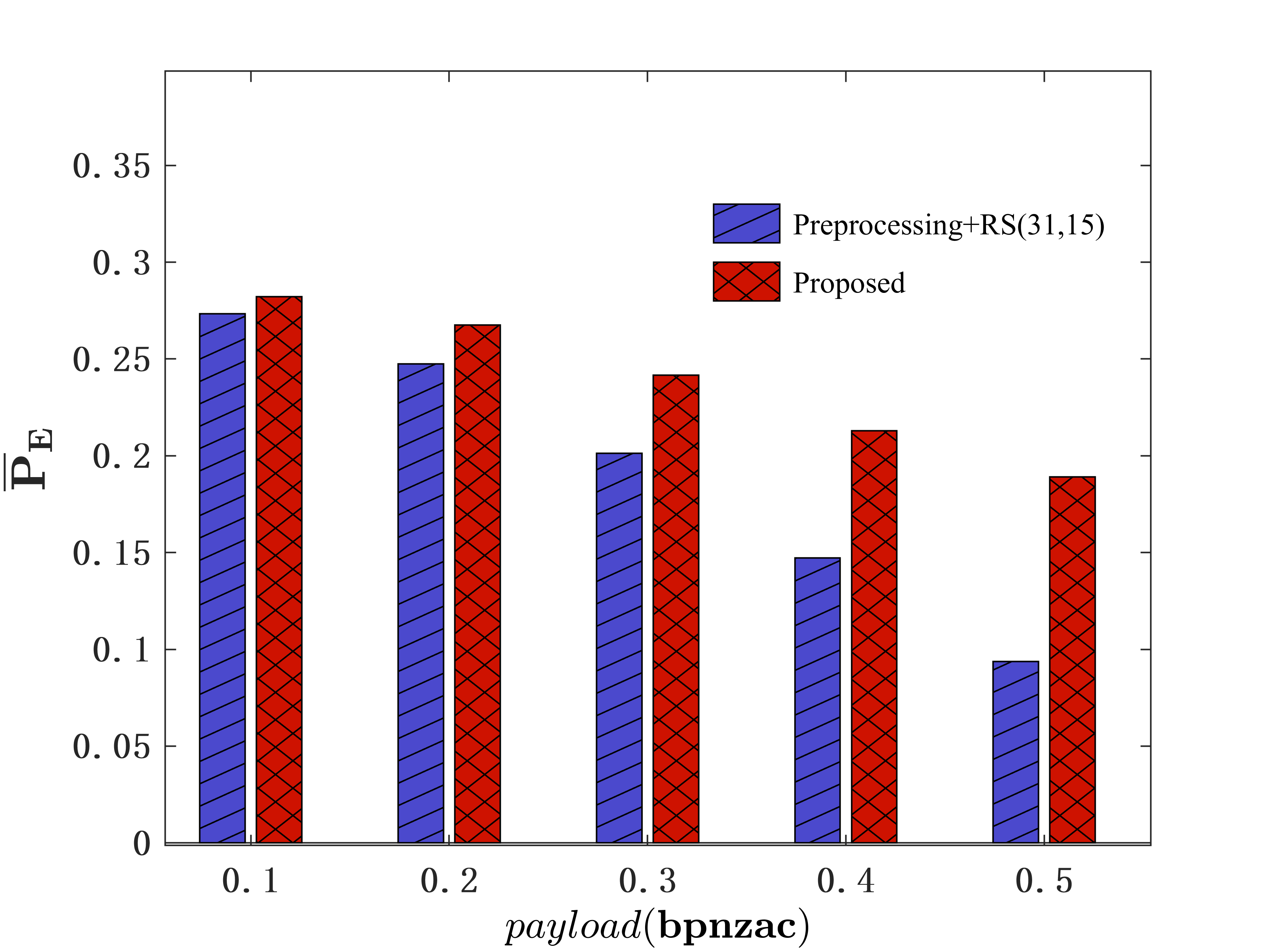}}
    \subfloat[DCTR]{
    \includegraphics[width=0.28\linewidth]{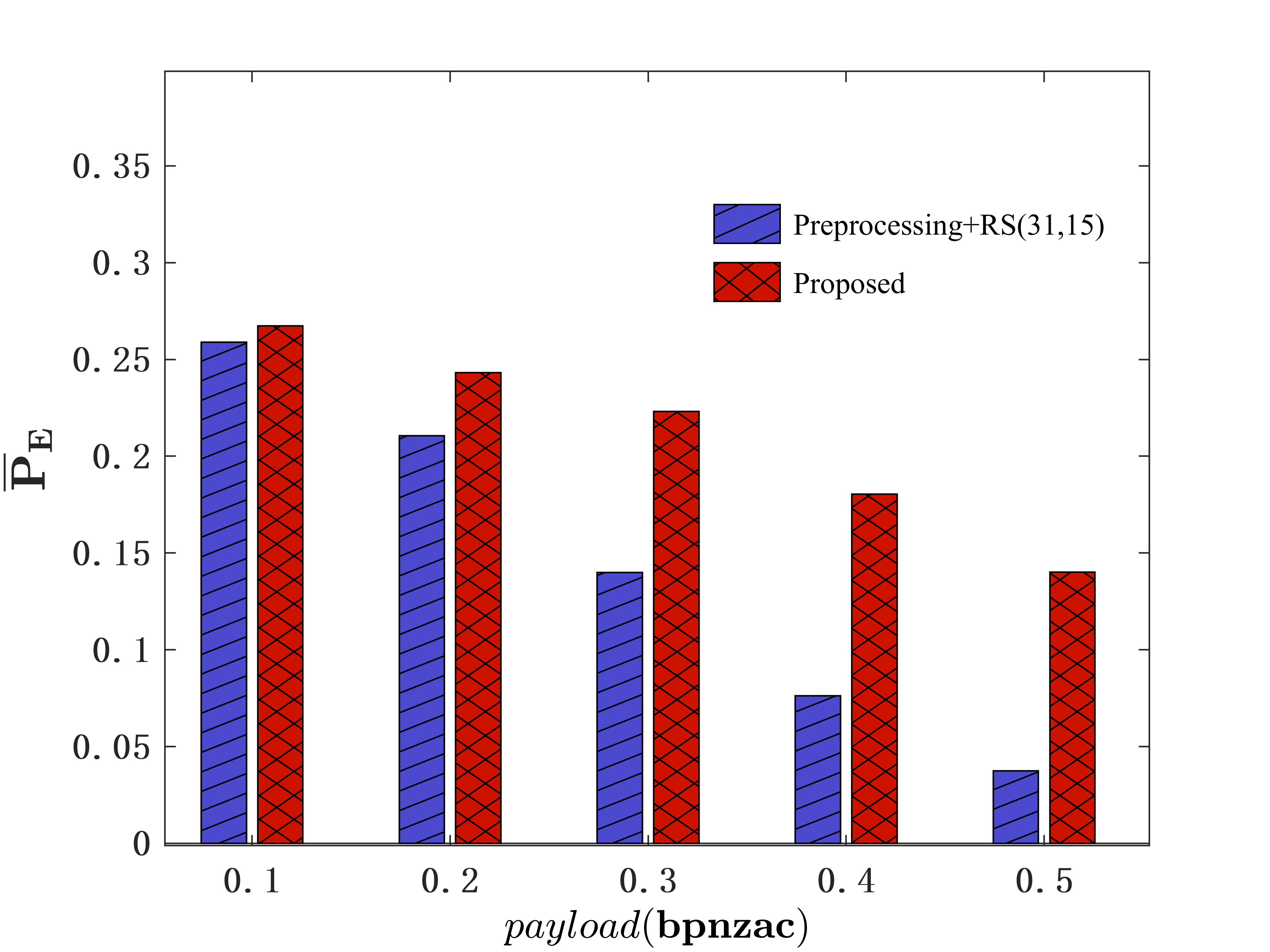}
    }
    \subfloat[Robustness]{
    \includegraphics[width=0.28\linewidth]{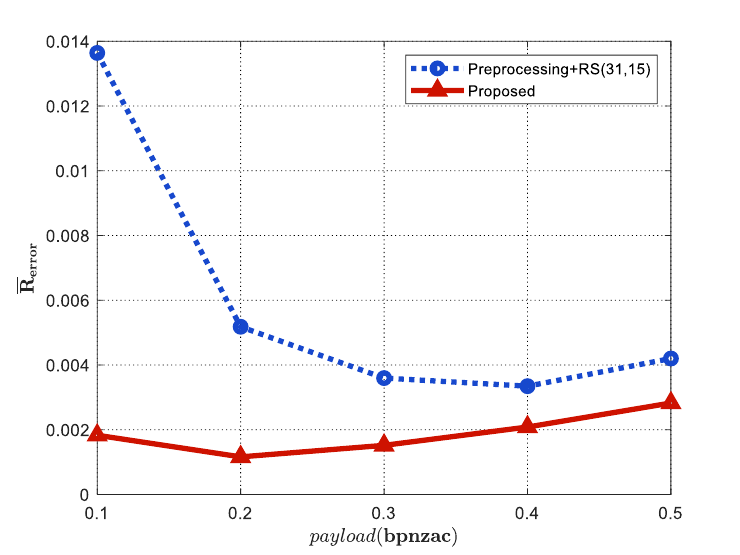}
    }
    \caption{Comparison of the performance about RS code at $Q_{channel}$ = 85.}
    \label{fig:ab2}
\end{figure*}
As the $Q_{channel}$ value increases and the quantization step \textit{\textbf{q}} decreases, the impact of JPEG recompression on image quality diminishes, thereby enhancing the robustness of steganographic techniques. Fig. \ref{fig:robustness}\subref{robust_95} illustrates the robustness of the proposed method for $Q_{channel}$ = 95, demonstrating its superiority over GMAS \citep{yu2020robust} and its outperformance of ROAST-ST \citep{10093140} at lower payload levels. Adaptive-GMAS \citep{duan2023robust} exhibits a decline in robustness with increasing embedding rates, leading to a higher average bit error rate. Overall, the proposed method offers significant advantages in practical applications due to its exceptional security while maintaining a certain level of robustness even at high embedding rates.

\subsection{Ablation Experiments about Preprocessing}

In this section, we demonstrate the effectiveness of our preprocessing method by comparing its performance with that of the ROAST \citep{10093140} preprocessing method.

\begin{table}[H]
\caption{Comparison of Image Quality After Preprocessing} 
\label{tab:eyes}
\centering 
\renewcommand\arraystretch{1.5} 
\scriptsize
\begin{tabular}{l p{2cm} p{2cm} p{2cm}} 
\toprule 
& \textbf{\textit{ROAST-OS} \citep{10093140}} & \textbf{\textit{ROAST-ST} \citep{10093140}} & \textbf{\textit{Proposed}} \\
\midrule 
\textbf{SSIM}  & 0.9823 & 0.9840 & \textbf{0.9885} \\
\bottomrule 
\end{tabular}
\end{table}

TABLE \ref{tab:eyes} compares the average SSIM for 10,000 images from BOSSbase v1.01 \citep{bas2011break}, utilizing both the proposed method and the ROAST (ROAST-ST and ROAST-OS) \citep{10093140}. Notably, the comparison involves images before and after preprocessing using different methods. It is evident that our preprocessing method produces images with superior quality. This can be attributed to minimizing modifications to these boundaries based on the actual overflow conditions of the images. By doing so, the proposed method preserves more texture and detail, resulting in better image quality compared to the ROAST (ROAST-OS and ROAST-ST).

Subsequently, we compare the performance of the two preprocessing methods (ROAST-ST and the proposed method) combined with adaptive error correction algorithms, as shown in Fig. \ref{fig:ab1}. It can be observed that the proposed method exhibits better anti-steganalysis performance. This is because the ROAST-ST preprocessing method is prone to making significant modifications to image pixel values, thereby affecting its anti-steganalysis performance. Furthermore, considering that overflow occurrences are more probable at spatial block boundaries and alterations to these boundaries can significantly influence empirical security, we augment the anti-steganalysis performance by minimizing adjustments to block boundaries. However, this reduction in modifications may potentially compromise the stability of quantized DCT coefficients, thereby affecting the robustness of the proposed method.

\subsection{Ablation Experiments about Adaptive Error Correction}

To demonstrate the improved performance in robustness and security offered by adaptive error correcting coding, this subsection compares the proposed method with the combination of the proposed preprocessing method and fixed RS(31,15) error correcting codes. Other configurations are consistent with the previous context. The precise comparisons of robustness and security performance are delineated in Fig. \ref{fig:ab2}.

As we can see, the steganographic technique with adaptive error correction codes outperforms the one with fixed error correction codes in terms of robustness and significantly enhances security. In particular, when the payload is 0.5, the proposed method improves the anti-steganalysis performance for DCTR \citep{holub2014low} by more than 10$\%$ compared to the method without adaptive error correction. This is because the adaptive adjustment of error correcting capabilities fully utilizes the inherent robustness of the image itself. If the image has strong robustness, a shorter error correcting code can be used to achieve the required error rate threshold while maintaining high security. In contrast, images with weaker robustness enhance robustness and prevent secret information corruption by increasing the length of the redundant error-correcting code. On the other hand, error-correcting codes with fixed lengths not only have lower coding efficiency but also rapidly decline in anti-steganalysis performance as the payload increases. Although adaptive error correction increases the time complexity of the algorithm, the performance improvement it brings is significant. Additionally, since the proposed method eliminates the need for embedding domain selection, its time complexity is much lower than that of the Adaptive-GMAS \citep{duan2023robust}.

\section{Conclusion}
\label{section:Conclusion}
This paper statistically analyzed the overflow characteristics of spatial blocks and found that the boundary regions of spatial blocks are more prone to overflow. Based on this finding, we propose an overflow preprocessing method aimed at reducing modifications in the boundaries of spatial blocks, thereby enhancing the security while maintaining the robustness of the algorithm. Additionally, we adaptively adjust the error correction coding capability to reduce redundancy in error correction coding, which effectively improves the robustness and security. The experimental results indicate that the proposed method outperforms existing approaches in anti-steganalysis performance while maintaining a high level of robustness. In the future, we will extend our work to color images and optimize the algorithm according to different application scenarios.

\printcredits

\bibliographystyle{cas-model2-names}

\bibliography{reference}


\end{document}